\documentclass[proof]{pasj01}
\usepackage{graphicx,natbib}

\begin{document} 

\title{VLBI observations of a flared optical quasar CGRaBS J0809+5341}


\author{Tao \textsc{An}\altaffilmark{1,2,3}}
\email{antao@shao.ac.cn}

\author{Yu-Zhu \textsc{Cui}\altaffilmark{1,2}}

\author{Zsolt \textsc{Paragi}\altaffilmark{4}}

\author{S\'{a}ndor \textsc{Frey}\altaffilmark{5}}

\author{Leonid~I. \textsc{Gurvits}\altaffilmark{4,6}}

\author{Krisztina~\'{E}. \textsc{Gab\'{a}nyi}\altaffilmark{5,7}}

\altaffiltext{1}{Shanghai Astronomical Observatory, Chinese Academy of Sciences, 200030 Shanghai, China}
\altaffiltext{2}{School of Electrical and Electronic Engineering, Shanghai Institute of Technology, 201418, Shanghai, China}
\altaffiltext{3}{Key Laboratory of Radio Astronomy, Chinese Academy of Sciences, 210008 Nanjing, China}
\altaffiltext{4}{Joint Institute for VLBI ERIC, Postbus 2, 7990 AA Dwingeloo, the Netherlands}
\altaffiltext{5}{F\"{O}MI Satellite Geodetic Observatory, PO Box 585, H-1592 Budapest, Hungary}
\altaffiltext{6}{Department of Astrodynamics and Space Missions, Delft University of Technology, Kluyverweg 1, 2629 HS Delft, the Netherlands}
\altaffiltext{7}{Konkoly Observatory, MTA Research Centre for Astronomy and Earth Sciences, PO Box 67, H-1525 Budapest, Hungary}



\maketitle

\begin{abstract}
A bright optical flare was detected in the high-redshift ($z=2.133$) quasar CGRaBS J0809+5341 on 2014 April 13. The absolute magnitude of the object reached $-30.0$ during the flare, making it the brightest one (in flaring stage) among all known quasars so far. The 15 GHz flux density of CGRaBS J0809+5341 monitored in the period from 2008 to 2016 also reached its peak at the same time. To reveal any structural change possibly associated with the flare in the innermost radio structure of the quasar, we conducted a pilot very long baseline interferometry (VLBI) observation of CGRaBS J0809+5341 using the European VLBI Network (EVN) at 5 GHz on 2014 November 18, about seven months after the prominent optical flare. Three epochs of follow-up KaVA (Korean VLBI Network and VLBI Exploration of Radio Astrometry Array) observations were carried out at 22 and 43 GHz frequencies from 2015 February 25 to June 4, with the intention of exploring a possibly emerging new radio jet component associated with the optical flare. However, these high-resolution VLBI observations revealed only the milliarcsecond-scale compact ``core'' that was known in the quasar from earlier VLBI images, and showed no sign of any extended jet structure. Neither the size, nor the flux density of the ``core'' changed considerably after the flare according to our VLBI monitoring. The results suggest that any putative radio ejecta associated with the major optical and radio flare could not yet be separated from the ``core'' component, or the newly-born jet was short-lived.
\end{abstract}

\section{Introduction}
\label{sect:intro}

Blazars are active galactic nuclei (AGN) with relativistic jets closely aligned with our line of sight according to the radio-loud AGN unification \citep{Urry95}. As a result, the Doppler-boosted relativistic jet emission dominates their non-thermal spectrum from the radio \citep{Blandford79} through optical \citep{Whiting01} to the $\gamma$-rays \citep{Ackermann11}. Phenomenological relations between optical flaring and radio properties in blazars have been investigated spanning a duration of more than four decades \citep[e.g.,][]{Hac72,Pom76,Babadzhanyants84,Tor94}. However, the physics of the inter-relating properties across the electromagnetic spectrum remains enigmatic.

Recent studies found a significant positive correlation between the optical nuclear luminosity and the radio flux density of the compact core in quasars, indicating that both the radio and optical emissions originate from the innermost part of the relativistically beamed pc-scale jets \citep{Ars10}. Correlations between the optical and $\gamma$-ray variability have also been found in blazars \citep{Hovatta14,Cohen14}, supporting the single-zone leptonic models in which the optical seed synchrotron photons are up-scattered by relativistic electrons to $\gamma$-ray energy bands via the inverse Compton process. As for possible correlations between $\gamma$-ray flares and the emergence of new superluminal VLBI components, \citet{Jor01} found a correspondence between these events in about half of the cases in their blazar sample, suggesting that the $\gamma$-ray emission is closely related to the relativistic jet. The physical mechanism producing the $\gamma$-ray flares is either synchrotron self-Compton or external Compton scattering of photons by relativistic electrons in the pc-scale regions of the jet. The location of the seed photon sources however may span two orders of magnitude in distance from the black hole, from the broad-line region (BLR, $\sim$0.1~pc), the molecular torus ($\sim$1$-$few pc), or the radio core ($\sim$10~pc) \citep{Dotson15}.

CGRaBS J0809+5341 (J0809+5341, hereafter) is a flat-spectrum radio quasar  \citep[a blazar;][]{Massaro09} at high redshift, $z = 2.133$ \citep{Healey08}. Recently it showed a bright flare in unfiltered optical observations \citep{Shumkov14,Balanutsa14}. The observations were made with the MASTER-Tunka auto-detection system on the nights of 2014 April 13 and April 19. The absolute magnitude of the flaring source was extremely high, $M=-30.0$ on April 13 and $M=-30.5$ on April 19, making it (during the short flaring period) possibly the brightest among all known quasars. In the subsequent observation on 2014 May 2, the source became significantly fainter, but it was still about 3 magnitudes brighter than in its quiescent state \citep{Wiersema14}. The source has recently been detected at high energies in the decaying part after the 2014 optical flare as well. It has been detected with the {\em Fermi} Large Area Telescope (LAT) \citep{Fermi15}; this observation indicated variability in the high-energy bands. It remained undetected during the first 2 years of {\em Fermi} operations but became active and continuously detected in the last year \citep{Paliya15}. In the same study, the source was also tracked in the X-rays with the {\em Nuclear Spectroscopic Telescope Array (NuSTAR)} and {\em Swift} satellites.

The strong optical flare and the recent high $\gamma$-ray state in J0809+5341 would be expected to cause a radio flux density outburst with the emergence of a new jet component 
\citep[e.g.,][]{Jor01,Mar08,Mar10,Ori13}. 
High-resolution VLBI imaging observations are essential to confirm this. Motivated by the discovery of the prominent optical flare, we have carried out a short exploratory VLBI observation of J0809+5341 with the European VLBI Network (EVN), with the aim of searching for possible structural changes and radio flux density variation associated with the event. The experiment was conducted on 2014 November 18, seven months after the optical flare. Then, we continued monitoring the source with the KaVA, a VLBI array combining the Korean VLBI Network (KVN) and the Japanese VLBI Exploration of Radio Astrometry (VERA) array, at frequencies of 22 and 43 GHz. The dual-frequency KaVA observations of J0809+5341 were conducted at three epochs (2015 February 25/26, 2015 April 2/3, and 2015 June 3/4), in order to trace possible structural variations, flux density variability, and the change of radio spectral index with time.

Here we report on the results of our VLBI observations of J0809+5341. The paper is organized as follows: Section \ref{sect:obs} describes the EVN and KaVA observations and the data analysis. Section \ref{sect:res} presents the results which are then discussed in Section \ref{sect:disc}. A summary of the current study is then presented in Section \ref{sect:sum}.

\section{Observations and data reduction}
\label{sect:obs}

\subsection{EVN observation and data reduction}
\label{subsect:evn}

\begin{figure*}
\centering
    \includegraphics[width=0.4\textwidth]{fig1a.ps}
    \includegraphics[width=0.4\textwidth]{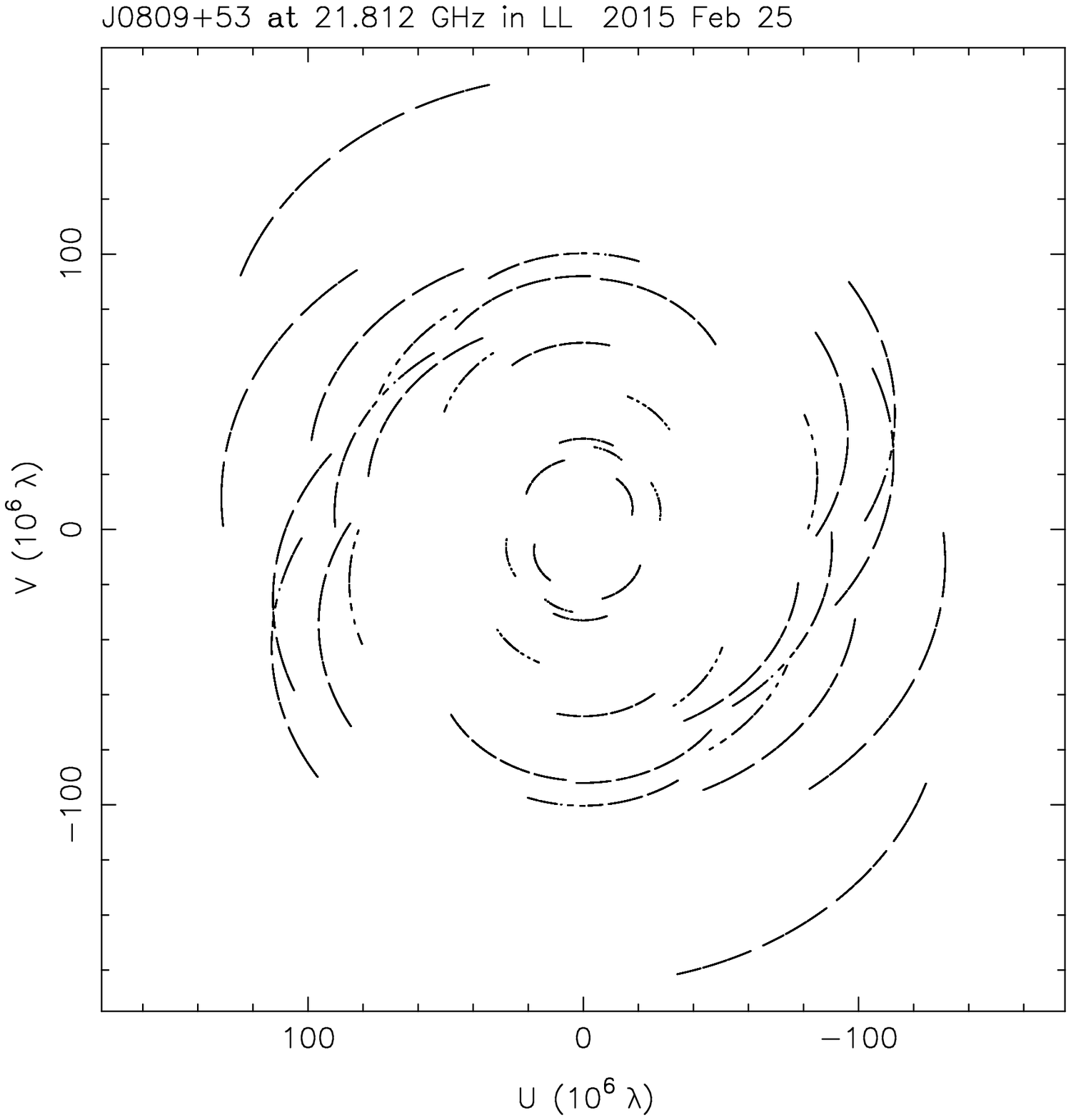}
	\caption{Left: The $(u,v)$-covergae of our initial e-EVN observations of J0809+5341 at 5~GHz. Right: An example of the $(u,v)$-coverage of KaVA observations of J0809+5341 at 22~GHz. The $u$ and $v$ components of the baseline vectors projected onto the plane perpendicular to the line of sight are expressed in the units of million wavelengths. As can be seen, the EVN and the KaVA observations have sampled similar spatial frequencies, but the distribution of the baselines is different. 
	}
	\label{fig:uvcov}
\end{figure*}

The observation of J0809+5341 was carried out with the EVN in electronic VLBI (e-VLBI) mode \citep{Szo08} at 5 GHz on 2014 November 18. Eight antennas participated in this experiment: Effelsberg (Germany), Jodrell Bank Mk2 (United Kingdom), Noto (Italy), Onsala (Sweden), Yebes (Spain), Toru\'n (Poland), Sheshan (China), and the Westerbork Synthesis Radio Telescope (WSRT, the Netherlands). The data observed at the telescopes were transmitted via wide-band optical fibre networks in real time to the EVN software correlator \citep[SFXC;][]{Kei15} at the Joint Institute for VLBI in Europe (JIVE), Dwingeloo, the Netherlands. The data were recorded in eight intermediate frequency channels (IFs) in both left and right circular polarizations at a maximum recording rate of 1024~Mbit~s$^{-1}$. The total bandwidth was 128 MHz. The observation was conducted in phase-reference mode \citep{BC95}. The telescopes nodded between the target and a nearby calibrator, J0809+5218, $1\fdg37$ away from the target. The target--reference duty cycle was 7 min long, with 4 min spent on J0809+5341. The total observing time was 2 h, and the effective on-target integration time was 1.2 h.

The post-correlated data were imported into the NRAO Astronomical Image Processing System ({\sc AIPS}), to calibrate the amplitudes and the fringe phases \citep{Diamond95}. After performing fringe-fitting \citep{Schwab83} for the phase-reference calibrator, the data were exported to the Caltech {\sc Difmap} package \citep{Shepherd94} for imaging and calibrating the residual phase errors. We performed a traditional hybrid mapping procedure consisting of several iterations of {\sc Clean}ing \citep{Hogbom74}, phase and amplitude self-calibration. The brightness distribution model of J0809+5218 was then used as input for a repeated fringe fitting in {\sc AIPS}, to account for the small residual phase errors caused by the non-pointlike structure of the calibrator. The derived gain solutions obtained for the calibrator were interpolated and applied to the target, J0809+5341. Then the visibility data were exported to {\sc Difmap} for imaging.

\begin{figure*}
\centering
\includegraphics[width=0.9\textwidth]{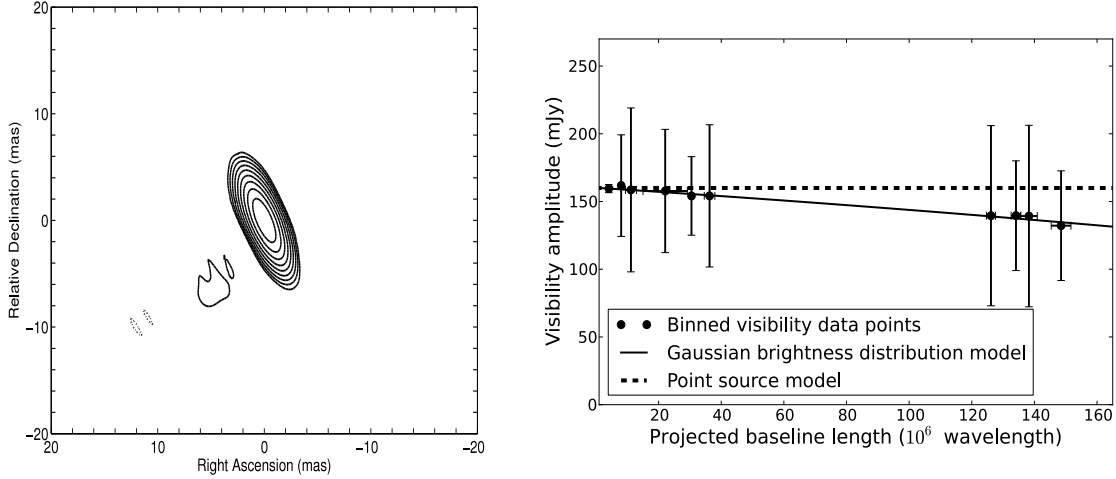}
\vspace{5mm}
\caption{Left: naturally weighted 5-GHz EVN image of J0809+5341. The image parameters (the lowest positive and negative contour level corresponding to the $\sim3 \sigma$ image noise, the peak brightness, and the restoring beam size and major axis position angle) are given in Table~\ref{tab:obs}. The positive contour levels increase by a factor of 2. 
Right: visibility amplitude versus projected baseline length. The visibilities with close {\it $(u,v)$} distances are binned together for illustration puproses only. The plot shows the weighted mean amplitudes and standard deviations. The solid curve indicates the best-fit circular Gaussian model.
}
\label{fig:evn}
\end{figure*}

As the target J0809+5341 itself was bright enough for fringe fitting, we repeated the procedure without using the complex gains from the phase-reference calibrator. For a comparison, the resulting image (shown in Fig.~\ref{fig:evn}) was similar to that obtained from the phase-referencing data reduction. The flux density uncertainty assumed as 10\% for the EVN originates from the errors of amplitude calibration based on the antenna gain curves and system temperature measurements.

\subsection{KaVA observations and data reduction}

The KVN \citep{Kim04} comprises of three radio telescopes of 21 m diameter, located at Seoul, Ulsan, and Jeju Island. The network is dedicated to high radio frequency (22, 43, 86, and 129~GHz) observations \citep{Han13}. The VERA array \citep{Kob03} consists of four telescopes of 20 m diameter located at Mizusawa, Iriki, Ogasawara, and Ishigaki-jima in Japan. It is dedicated to high-precision VLBI astrometric measurements. VERA antennas have been installed with dual-beam and 22/43~GHz receiver systems for efficient phase-reference VLBI observations.
The combination of KVN and VERA yields a new, powerful VLBI facility called KaVA (KVN and VERA array). KaVA thus consists of seven radio telescopes with the longest and shortest interferometric baselines of 2270 km and 305 km, respectively. The array is remarkable for its evenly distributed $(u,v)$ spacings (Fig.~\ref{fig:uvcov}). The imaging capability of KaVA for extended radio structures in bright AGN has been demonstrated by \citet{Nii14}.

\begin{figure*}
\centering
\includegraphics[width=1.0\textwidth]{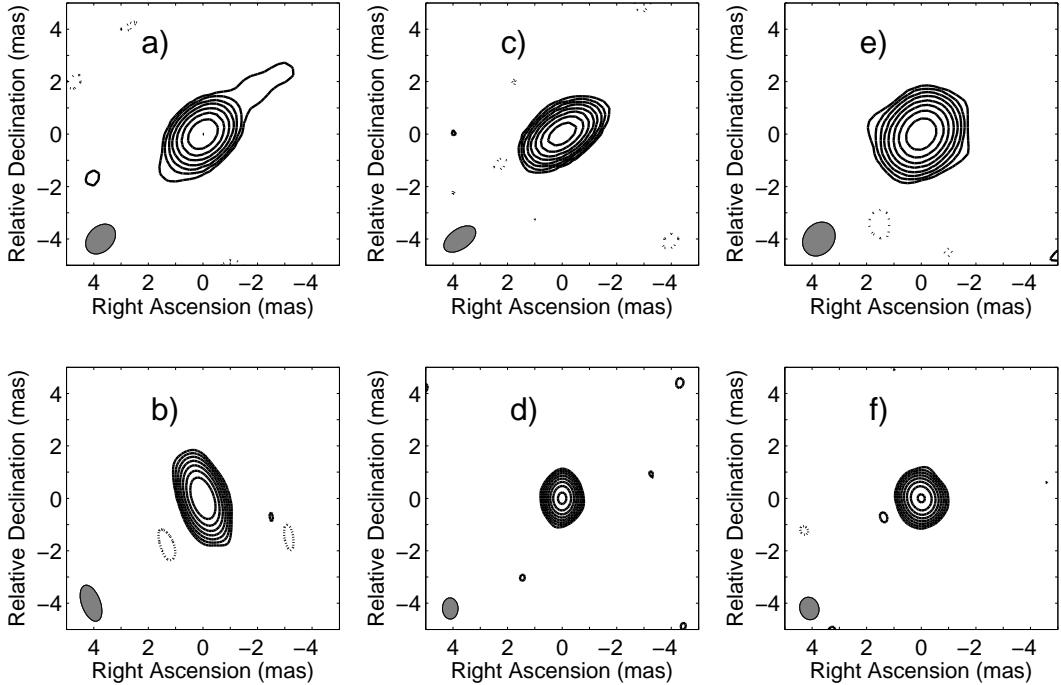}
\caption{Naturally weighted KaVA images of J0809+5341 at 22~GHz (top) and 43~GHz (bottom). The observing epochs and image parameters (the lowest positive and negative contour level corresponding to the $\sim3 \sigma$ image noise, the peak brightness, and the restoring beam size and major axis position angle) are presented in Table~\ref{tab:obs}. The positive contour levels increase by a factor of 2. The ellipse in the bottom-left corner indicates the restoring beam.}
\label{fig:KaVA}
\end{figure*}

Our KaVA observations of J0809+5341 were made at two frequencies, 22 and 43~GHz. Since the aim of the project was to detect changes in the putative jet structure, the observations were scheduled in three sessions separated by about 5 and 8 weeks: 2015 February 25/26, 2015 April 2/3, and 2015 June 3/4. All seven KaVA antennas participated in these experiments. Occasionally, for periods of time, the measured system temperatures at certain individual antennas significantly exceeded the nominal values. Affected data were omitted from the subsequent analysis. The KaVA data were recorded in 16 IFs in left-hand circular polarization with two-bit quantization, at a rate of 1024~Mbit~s$^{-1}$. The total bandwidth was 256~MHz. The observing time at each frequency on each date was 4~h. Apart from five 4-min scans on a bright calibrator (4C\,39.25) which were used for fringe finding, the remaining time was spent on J0809+5341. The effective observing time spent on the target was nearly 3.7~h at both frequencies in each session.
 
The data at the VERA and the KVN telescopes were recorded on magnetic tapes and disks, respectively. After completing the experiments, the tapes and disks were shipped from the stations to the Korea--Japan Correlation Center (KJCC) located in Daejeon, Korea \citep{Lee14,2015JKAS...48..125L}, where the correlation was performed. The correlated interferometric visibility data were imported into {\sc AIPS} where the amplitudes and phases were calibrated. The data reduction followed the standard procedure described in Section~\ref{subsect:evn} \citep[see also][]{Lee15} with direct fringe-fitting to the J0809+5341 data. An amplitude calibration uncertainty of 15\% is assumed for KaVA \citep[e.g.,][]{Nii14}. The imaging with {\sc Clean} and phase self-calibration cycles was performed in {\sc Difmap}. One iteration of amplitude and phase self-calibration was done to improve the dynamic range of the image, but no more amplitude self-calibration step was performed, to avoid the inclusion of a scaling error of the visibility amplitude.

\section{Results}
\label{sect:res}

\subsection{The VLBI structure}
\label{size}

Figure~\ref{fig:evn} shows the 5-GHz EVN image of J0809+5341, characterized by a single compact component, apparently the radio ``core''. It is conventionally interpreted as the inner section of the synchrotron-emitting jet that becomes optically thick at the given observing frequency \citep{Blandford79}. This image resulted from fringe-fitting to the J0809+5341 data. The position of the emission peak in the phase-referenced image (not reproduced here) is right ascension $08^{\rm h} 09^{\rm m} 41\fs7330$ and declination $+53\degree 41\arcmin 25\farcs093$ (J2000), which is in good agreement with that derived from the analysis of the data from the VLBI calibrator database \footnote{Data from http://astrogeo.org maintained by L. Petrov, solution rfc\_2014d}.

\begin{table*}
  \tbl{Parameters of the VLBI images.}{%
  \begin{tabular}{ccccccccc}
  \hline
Epoch & Array & $\nu$     & Restoring beam  FWHM      & Lowest contours    & Peak brightness & Figure \\
      &       & (GHz)     & (major axis $\times$ minor axis, PA) & (mJy beam$^{-1}$)  &   (Jy beam$^{-1}$)  \\
\hline
2014 Nov 18  & EVN   & 5.0   & 4.65 mas$\times$1.37 mas, $23\fdg3$  & 0.3 & 0.158 & \ref{fig:evn}   \\
2015 Feb 25  & KaVA  & 21.8  & 1.27 mas$\times$0.93 mas, $-41\fdg4$ & 1.2 & 0.154 & \ref{fig:KaVA}a \\
2015 Feb 26  & KaVA  & 43.5  & 1.43 mas$\times$0.69 mas, $18\fdg7$  & 0.7 & 0.114 & \ref{fig:KaVA}b \\
2015 Apr 02  & KaVA  & 21.8  & 1.34 mas$\times$0.75 mas, $-53\fdg7$ & 0.7 & 0.154 & \ref{fig:KaVA}c \\
2015 Apr 03  & KaVA  & 43.5  & 0.80 mas$\times$0.57 mas, $-1\fdg1$  & 0.8 & 0.134 & \ref{fig:KaVA}d \\
2015 Jun 03  & KaVA  & 21.8  & 1.36 mas$\times$1.11 mas, $-33\fdg5$ & 0.7 & 0.171 & \ref{fig:KaVA}e \\
2015 Jun 04  & KaVA  & 43.5  & 0.86 mas$\times$0.70 mas, $13\fdg1$  & 0.9 & 0.126 & \ref{fig:KaVA}f \\
\hline
\end{tabular}}\label{tab:obs}
\begin{tabnote}
\end{tabnote}
\end{table*}

Figure~\ref{fig:KaVA} shows the 22 and 43 GHz emission structure observed with KaVA at three epochs. The images are made with natural weighting. The highest resolution is 0.57~mas at 43~GHz and 0.75~mas at 22~GHz, corresponding to a projected linear size of 4.7~pc and 6.2~pc, respectively\footnote{We assume a flat cosmological model with $H_{\rm 0}$=70~km~s$^{-1}$~Mpc$^{-1}$, $\Omega_{\rm m}$=0.3, and $\Omega_{\Lambda}=$0.7, which gives a scaling parameter of 8.3~pc\,mas$^{-1}$.}. The image parameters are listed in Table~\ref{tab:obs}. The source does not show any obvious extended radio structure on these scales down to the brightness level of $0.3$~mJy\,beam$^{-1}$.

For a simple characterization of the brightness distribution of the compact ``core'', we fitted circular Gaussian model components to the self-calibrated VLBI visibility data in {\sc Difmap}. The results of the model fitting are presented in Table~\ref{tab:mf}. The fit to the 5 GHz EVN data gave a $159 \pm 16$~mJy flux density and $0.31 \pm 0.06$~mas de-convolved angular size (full width at half-maximum, FWHM). 
The error in the core size was estimated by using Monte Carlo simulations and allowing for a 10\% variation in the visibility amplitudes. 
The right panel of Fig.~\ref{fig:evn} shows the visibility amplitudes as a function of the projected baseline length, demonstrating that the source
is slightly resolved. The 0.31-mas diameter circular Gaussian model fitted in {\sc Difmap} using all individual visibility data points is also indicated as solid line, as well as a point source model with 159~mJy flux density shown as dashed line.

The fitted core size from the KaVA data ranges from 0.06 to 0.10~mas (Table~\ref{tab:mf}). While the resolution of the EVN and KaVA
observations are similar (cf. Fig.\ref{fig:uvcov}), the core size derived from the KaVA data is about three times smaller than that from the EVN data.
This is not surprising as at the 4--8 times higher observing frequencies, as the KaVA is getting emission from the inner-most region of the jet, and the core size is expected to scale with $\nu^{-1}$ \citep{Blandford79}. We therefore note that the highest frequency KaVA data may actually be consistent with a point source and the fitted sizes are upper limits. 

Following the same procedure, we also analyzed the archival 5 GHz VLBA data from the VIPS project \citep{Helmboldt07}, and 2.3/8.4 GHz data from the VLBA Calibrator Survey \citep[VCS1,][]{Beasley02}. The fitted model parameters are listed together with those derived from our EVN and KaVA observations in Table~\ref{tab:mf}.  

\begin{table*}
  \tbl{The results of circular Gaussian model fitting to VLBI data of J0809+5341 and the source physical parameters derived.}{%
  \begin{tabular}{ccccccc}
  \hline
Epoch & $\nu$ & $S$   & $\theta$ & $T_{\rm b}$   & $\delta$ & Reference \\
      & (GHz) & (mJy) & (mas)    & ($10^{11}$~K) &          &           \\
\hline
1994 Aug 12  & 2.3   & 171$\pm$17 & 1.17$\pm$0.06 & 0.9$\pm$0.1 &  1.8$\pm$0.2 & \citet{Beasley02} \\
1994 Aug 12  & 8.4   & 143$\pm$14 & 0.24$\pm$0.02 & 1.3$\pm$0.2 &  2.6$\pm$0.3 & \citet{Beasley02} \\
2006 May 31  & 4.8   & 184$\pm$18 & 0.24$\pm$0.03 & 5.3$\pm$1.4 & 10.6$\pm$2.4 & \citet{Helmboldt07} \\
2014 Nov 18  & 5.0   & 159$\pm$16 & 0.31$\pm$0.07 & 2.5$\pm$0.6 &  5.0$\pm$1.1 & this paper \\
2015 Feb 25  & 21.8  & 154$\pm$23 & 0.06$\pm$0.01 & 1.2$\pm$0.3 &  2.4$\pm$0.5 & this paper \\
2015 Feb 26  & 43.5  & 116$\pm$17 & 0.10$\pm$0.02 & $>$0.20$\pm$0.05 & $>$0.4$\pm$0.1  & this paper \\
2015 Apr 02  & 21.8  & 155$\pm$23 & 0.07$\pm$0.02 & 2.2$\pm$0.7 &  4.4$\pm$1.4 & this paper \\
2015 Apr 03  & 43.5  & 135$\pm$20 & 0.07$\pm$0.03 & $>$0.6$\pm$0.3 & $>$1.2$\pm$0.6 & this paper \\
2015 Jun 03  & 21.8  & 170$\pm$25 & 0.10$\pm$0.02 & 1.3$\pm$0.3 &  2.6$\pm$0.7 & this paper \\
2015 Jun 04  & 43.5  & 127$\pm$19 & 0.09$\pm$0.03 & $>$0.3$\pm$0.1 & $>$0.6$\pm$0.2 & this paper \\
\hline
\end{tabular}}\label{tab:mf}
\begin{tabnote}
\end{tabnote}
\end{table*}

\subsection{Flux density variability and radio spectrum}
\label{variability}

The modelfits to the VLBI data (Table~\ref{tab:mf}) do not exclude the variability of the compact radio component. However, these flux densities are measured at one (2.3 and 8.4~GHz), two (5~GHz) or three (22 and 43~GHz) epochs only, and the assumed 10-15\% amplitude calibration uncertainties make the different values consistent with each other within the uncertainties.

The historical total flux density measurements of J0809+5341, collected in the NASA/IPAC Extragalactic Database\footnote{http://ned.ipac.caltech.edu/} (NED) hint at variability. The single-dish Green Bank observation at 1.4~GHz made by \citet{WB92} yields a 180~mJy flux density, while at 5~GHz in 1987 October, it was 197~mJy \citep{BWE91,GC91}. The integrated 1.4~GHz flux density in the U.S. National Radio Astronomy Observatory (NRAO) VLA Sky Survey \cite[[NVSS, ][]{Condon98}
is $140.4 \pm 4.2$~mJy. This made use of the most compact D-array configuration of the VLA (providing an angular resolution of 45\arcsec).
The flux density measured by the B-array configuration in the VLA Faint Images of the Radio Sky at Twenty-Centimeters (FIRST) survey \citep{Becker95} is $131.7 \pm 0.2$~mJy. The difference between the VLA interferometer and Green Bank single-dish measurements is $\sim$40~mJy (nearly 30\%). This could in principle be a resolution effect, but the source is known to be compact on arcsec scales and below, as is obvious from, among others, our own EVN and KaVA observations. Thus the flux density difference cannot be attributed to arcmin-scale extended emission, but rather the time variability of the compact component. We note that J0809+5341 appears as an unresolved optical object in the Sloan Digital Sky Survey \citep[SDSS,][]{Aba09} image on arcsec scale.

\begin{figure}
\centering
\includegraphics[width=0.5\textwidth]{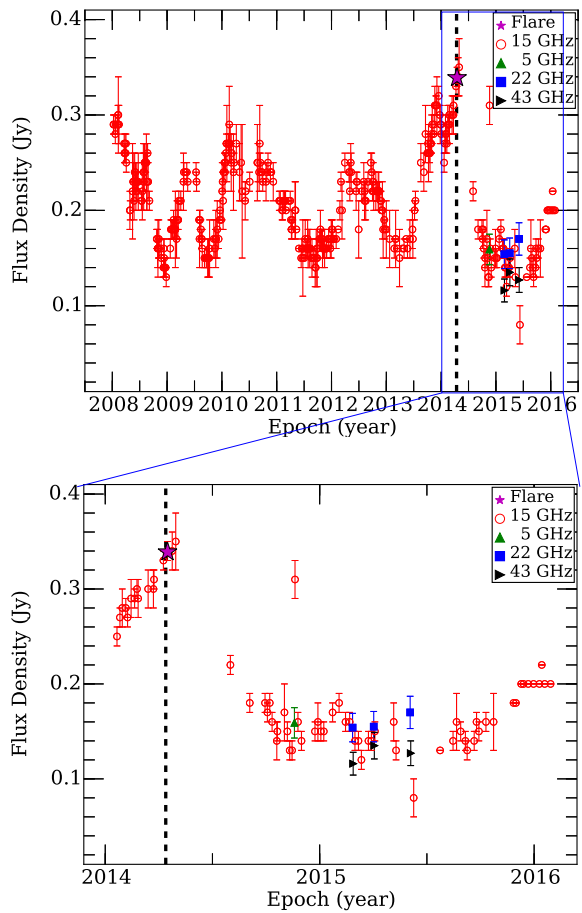}
\vspace{5mm}
\caption{The radio light curve of J0809+5341 observed in the OVRO monitoring programme \citep{Richards11} at 15~GHz, from 2008 to 2016 (circles). The date of the major optical flare in 2014 is indicated with a dashed vertical line. Our VLBI ``core'' flux density data points are plotted with different symbols: 5-GHz EVN (up triangle), 22-GHz KaVA (square) and 43-GHz KaVA (right triangle). 
}
\label{fig:lc}
\end{figure}

J0809+5341 is being monitored at 15~GHz with the Owens Valley Radio Observatory (OVRO) 40-m radio telescope\footnote{http://www.astro.caltech.edu/ovroblazars/} as part of a large blazar sample \citep{Richards11}. Figure~\ref{fig:lc} shows the OVRO light curve of J0809+5341 covering a period of more than 8 years, from 2008 to 2016. For more details on the OVRO blazar monitoring program and data reduction, we refer to \citet{Richards11}. According to Fig.~\ref{fig:lc}, J0809+5341 is highly variable with flux densities ranging from $\sim$120 to 360~mJy. The light curve is characterized by a number of subsequent flares. The large flares last for about 2 years. Each large flare appears to show a double peak with a separation of several months. Starting from 2013 February, the source went into a flaring phase. 
The prominent optical flare discovered in 2014 April \citep{Shumkov14,Balanutsa14} coincides with the peak of radio emission, but since the source has reached similar brightness at two other occasions during the 8-yr monitoring period, this might be just  coincidental.

Overall, the source J0809+5341 shows a flat radio spectrum with a spectral index $\alpha=0.04$ ($S \propto \nu^\alpha$, where $\nu$ is the frequency and $S$ is the flux density), as derived from the non-simultaneous data with the inhomogeneous resolutions available in NED. We can also use our nearly simultaneous 22 and 43 GHz KaVA data to estimate the spectral index of the compact core. That results in $\alpha_{43}^{22} = -0.4 \pm 0.2$ in the first and third epochs, and $\alpha_{43}^{22} = -0.2 \pm 0.1$ in the second epoch. 

\section{Discussion}
\label{sect:disc}

\subsection{Source variability}

As was described in Sect.~\ref{variability}, the source has a flat radio spectrum and shows strong variability in the radio. This is consistent with the blazar classification of J0809+5341. Following \citet{Ars10}, we calculated the radio-loudness parameter $R$, defined as the ratio of the radio flux density at 5~GHz to the nuclear optical flux density at 4400~\AA{} \citep{Kel89}. For a source at $z = 2.133$, the above bands used for calculating $R$ correspond to 1.6~GHz and 13800~\AA{} in the observer's frame. J0809+5341 shows a practically flat radio spectrum, allowing us to assume a 1.6~GHz flux density of 160~mJy in the quiescent state, and 320~mJy at the flare peak. In the flaring state, the MASTER OT observatory detected the unfiltered R-band magnitude of 16.2 \citep{Shumkov14,Balanutsa14}. Compared to the historical R-band data, it is 3.4 mag lower. When converting to flux density at 13800~\AA, it corresponds to about 4.5~mJy. Then, the radio-loudness of J0809+5341 is $R = 70$ during the 2014 optical flare. As a comparison, $R = 1800$ is estimated during quiescence, reinforcing the classification of the object as a radio-loud quasar.

Optical flaring of blazars have been studied for over five decades \citep[e.g.,][]{GK65,Pol79,AS80}, and correlations between optical and radio flares have been detected in some cases, e.g., in the prominent radio AGN AO~0235+164 and 3C~345 \citep{BD80,Babadzhanyants84}. Long-term multi-band monitoring of a sample of blazars shows a tight correlation between the radio and optical luminosities \citep{Ars10,Wie15}. At the moment, it is not clear yet whether a similar correlation exists for J0809+5341, but we note that the major optical flare in 2014 took place at the same time when the 15 GHz radio flux density reached its the peak in April (Fig.~\ref{fig:lc}). Our VLBI observations were performed several months later, when the total radio flux density had already dropped considerably. The VLBI flux densities at various frequencies and epochs reported in this paper show a good consistency with the total flux density of the 15 GHz OVRO light curve (Fig.~\ref{fig:lc}), indicating that the total flux density is dominated by the compact VLBI core, and the core has a flat radio spectrum during this relatively quiet stage.

Recently, \citet{Paliya15} reported the first detection of J0809+5341 in the X-ray and $\gamma$-ray bands. This increase in high-energy emission is coincidental and likely associated with the giant optical flare. As mentioned earlier, the seed photon sources for inverse Compton scattering may originate in the BLR, the molecular torus, or the radio core. \citet{Paliya15} found that the $\gamma$-ray properties are consistent with an emission region outside of the BLR. In this case the flaring radio emission is expected to be completely synchrotron self-absorbed, and the observed maximum in radio flux density is likely a chance coincidence. When the shocked ejecta travel along the jet, we expect to see an increase of radio emission as it becomes transparent, first at the highest frequencies, 
as predicted by the shock-in-jet model \citep{MG85,Val92}. 
The fact that there has been no increase at 43~GHz in our monitoring implies that the flaring radio emission was either very short-lived, or the shocked ejecta has not propagated yet to the optically thin region.

Alternatively, the optical flare and the radio outburst, as well as the increase in the high-energy flux are physically related. Most blazar outbursts are known to occur at pc-scale distances from the central engine, around the radio core region \citep[see e.g.][for a review]{Mar13}. This can be confirmed by long-term monitoring observations with dense time sampling, from radio to $\gamma$-rays, supplemented with high-resolution VLBI monitoring in the radio. Such programmes are being undertaken for some of the most prominent blazars \citep[e.g.,][]{Mar08,Mar10,Agudo11a,Agudo11b,Ori13,Jor13} but not for J0809+5341. However, it is also possible that no new jet component was associated with the flare of J0809+5341, as, e.g., found for the blazar Mrk 421 by \citet{PE05}. This would suggest that the jet rapidly loses its kinetic energy and does not reach the region that can be imaged with the resolution offered by VLBI.

\subsection{The brightness temperature and the implications for the Doppler-boosting factor}

Based on the VLBI-measured flux density and source size presented in Sect.~\ref{size}, we calculated the apparent brightness temperature of J0809+5341 using the following equation \citep{K&O88}:
\begin{equation}
T_{\rm b} = 1.22 \times 10^{12} \frac{S_{\rm core}}{\nu^2 \theta^2} (1+z) ,
\end{equation}
where $T_{\rm b}$ is the brightness temperature in Kelvin, $S_{\rm core}$ [Jy] is the flux density of the ``core'' at the observing frequency $\nu$ [GHz], $\theta$ [mas] is the FWHM size of the best-fit circular Gaussian model. The redshift is $z= 2.133$ \citep{Healey08}. The calculated brightness temperatures for the different VLBI experiments are listed in Table~\ref{tab:mf}. The brightness temperatures are in the range of $(0.2 - 5.3) \times 10^{11}$\,K. These values are typical for most other radio-loud quasars observed with VLBI at around $z = 3$ \citep{Gurvits92,Gurvits94,Frey97,Par99}. 

The $T_{\rm b}$ values at 43~GHz appear consistently smaller than those measured at lower frequencies (Table~\ref{tab:mf}). This phenomenon has also been found in previous high-frequency VLBI surveys \citep{Lee08}. The difference derived from the statistical investigation of large samples is not simply due to source variability or other observing effects. A possible reason might be related to non-zero gradients in the physical conditions in the jet flow, and high-frequency (43- and 86-GHz) VLBI observations probe the optically thin region where $T_{\rm b}$ is intrinsically lower \citep{Lee08}. On the other hand, as we pointed out in Sect. 3.1, the core is not completely resolved, and in our case the fitted component size represents an upper limit at the highest frequencies. This means that the calculated $T_{\rm b}$ values at 43~GHz are in fact lower limits. Therefore we cannot independently confirm the decrease of $T_{\rm b}$ with frequency for J0809+5341. 

The brightness temperature of blazars is amplified by the Doppler boosting effect as the approaching jets are oriented close to the line of sight. Usually, the equipartition brightness temperature $T_{\rm b,eq} \simeq 5 \times 10^{10}$\,K \citep{Readhead94} is considered to be a reasonable estimate of the intrinsic value $T_{\rm b,int}$. The Doppler boosting factor is thus derived from the observed brightness temperature as $\delta = T_{\rm b}/T_{\rm b,eq}$. The estimated Doppler factors (lower limits in cases where the $T_{\rm b}$ values are lower limits as well) for J0809+5341 listed in Table~\ref{tab:mf} range from at least 0.4 to 10.6. 
The observations of $\delta<1$ at some epochs might indicate a
non-stationary flow of plasma resulting in both deviations from
equipartition as well as projection effects of a curved plasma flow
trajectory. However we note that the $\delta$ values somewhat below unity in Table 2
are all estimated at 43 GHz and, as discussed above, are lower limits
because the source is unresolved at this frequency. Therefore the lower values
are not inconsistent with the presence of Doppler boosting in the
jet.

\subsection{Comparison to jet parameters derived from high-energy observations}

The spectral energy distribution (SED) of the source during the flare was fitted by \citet{Paliya15} with a synchrotron self-Compton model, confirming that J0809+5341 is a powerful blazar. \citet{Paliya15} note however that the high-energy properties of J0809+5341 are reminiscent of low-redshift blazars rather than high-redshift ones. Its optical spectrum is dominated by synchrotron emission from the jet rather than an extremely luminous accretion disk; its $\gamma$-ray spectrum is flat rather than steep; and, it hosts a relatively low-mass black hole ($10^{8.4} M_{\odot}$ )  \citep[cf.][]{Ghisellini11,Ghisellini13}. Our VLBI result reveals a relativistic jet with a moderate Doppler boosting factor, consistent with typical blazar radio properties in general. From the SED, \citet{Paliya15} estimate a bulk Lorentz factor of $\Gamma=20$ in the jet, and suggest that the jet becomes radiatively efficient during the flare.  

Assuming the jet parameters obtained from SED fitting by \citet{Paliya15}, we independently estimate the Doppler factor $\delta$, following, e.g., \citet{Urry95}:
\begin{equation}
\delta=[\Gamma (1 - \beta \cos \vartheta)]^{-1},
\end{equation}
where the bulk velocity measured in the units of the speed of light $c$ is
\begin{equation}
\beta=(1 - \Gamma^{-2})^{\frac{1}{2}}.
\end{equation}
Substituting $\Gamma=20$ and the jet viewing angle $\vartheta=3.0\degree$ \citep{Paliya15}, we get $\delta=19.1$. It is higher compared to the values we derived from VLBI data (Table~\ref{tab:mf}). A possible reason is that we overestimate the intrinsic brightness temperature $T_{\rm b,int}$ by a factor of $\sim$2 by adopting the equipartition value $T_{\rm b,int}$  \citep[cf.][]{Homan06}, and thus underestimate the Doppler factor by the same factor.

In the relativistic beaming model applied to the parameters of J0809+5341, the observed transverse speed of a radio-emitting blob in the jet, expressed in the units of $c$ is
\begin{equation}
\beta_{\rm app}=\frac{\beta \sin \vartheta}{1 - \beta \cos \vartheta} = 19.95.
\end{equation}
Assuming the jet model for J0809+5341 proposed by \citet{Paliya15}, using $\delta=19.1$ for the Dopler factor, we can estimate the expected apparent proper motion $\mu$ of a putative newly-ejected superluminal jet component possibly associated with the optical flare (and the coincident radio and high-energy emission peak) in 2014 April, following \citet{Bach05}:
\begin{equation}
\mu=\beta_{\rm app} c (1+z) D_{\rm L}^{-1}.
\end{equation}
Here $D_{\rm L}=16809.4$~Mpc \citep{Wright06}, and thus $\mu=0.23$~mas~yr$^{-1}$. This slow apparent proper motion is consistent with our results, in particular with the fact that we did not detect any sign of a new jet component in our follow-up VLBI observations within 1.1~yr after the flare, with angular resolutions $\gtrsim 0.6$~mas (see Table~\ref{tab:obs}). If there was an emerging blob in the jet, then it was still blended with the ``core''. Another possibility is that the flare did not generate a jet component. Follow-up VLBI imaging over a sufficiently long time interval may eventually reveal a jet ejection, unless the blob is expanding and fading too rapidly to be detected several years after the flare.

\section{Summary}
\label{sect:sum}

We presented 5-GHz EVN and 22/43-GHz KaVA imaging results of J0809+5341 observed 7 months to 1.1~yr after the detection of its largest optical flare in 2014 April. Our high-resolution radio images (Figs.~\ref{fig:evn} and \ref{fig:KaVA}) reveal a compact unresolved core with the flux density of $\sim$160~mJy. Frequent single-dish monitoring observations at 15~GHz with the OVRO 40-m radio telescope (Fig.~\ref{fig:lc}) in the period between 2008 and 2016 showed that source flux density was changing within about a factor of two compared to its quiescent level.  
Brightness temperature and Doppler boosting factor estimated for J0809+5341 from the VLBI data are consistent with the presence of a relativistically beamed blazar jet. This conclusion, and the blazar identification of J0809+5341 is reinforced by recent X-ray and $\gamma$-ray data \citep{Paliya15} which also provides evidence for relativistic jet beaming. If there was any jet ejection associated with the major flare, the estimated slow apparent proper motion ($\mu=0.23$~mas~yr$^{-1}$) of the blob explains why no significant structural change of the compact radio source could be detected within only 1.1~yr after the event.

\begin{ack}
This work was supported by the SKA pre-construction funding from the China Ministry of Science and Technology under grant No. 2013CB837900. The research leading to these results has received funding from the European Commission Seventh Framework Programme (FP/2007-2013) under grant agreement No. 283393 (RadioNet3). Funding was received from the Hungarian National Research, Development and Innovation Office (OTKA NN110333), and the China--Hungary Collaboration and Exchange Programme by the International Cooperation Bureau of the Chinese Academy of Sciences.

We thank the referee for the constructive comments. 
This work is based [in part] on observations made with the KaVA, which is operated by the the Korea Astronomy and Space Science Institute and the National Astronomical Observatory of Japan.
We are grateful to all staff members and students at the KVN and VERA who helped to operate the array and to correlate the data. 
We are grateful to the chairman of the EVN Program Committee, Tom Muxlow, for accepting our short e-EVN proposal. The EVN is a joint facility of independent European, African, Asian, and North American radio astronomy institutes. Scientific results from data presented in this publication are derived from the following EVN project code: RSC02. 
We are grateful to the OVRO group for making their total flux density monitoring data available and Talvikki Hovatta for providing the light curve data. 
The research has made use of the Astrogeo Center database of brightness distributions, correlated flux densities, and images of compact radio sources produced with VLBI.
\end{ack}

\end{document}